\documentclass[journal=jacsnano,manuscript=article]{achemso}

\usepackage[version=3]{mhchem} 
\usepackage[usenames,dvipsnames]{xcolor}
\usepackage{mathptmx}
\usepackage{graphicx} 
\usepackage{hyperref}
\usepackage{nameref}

\usepackage[normalem]{ulem}

\author{Sergiu Ruta}
\email{sergiu.ruta@shu.ac.uk}
\altaffiliation{These authors contributed equally}
\affiliation{College of Business, Technology and Engineering, Sheffield Hallam University, United Kingdom}

\author{Yilian Fernández-Afonso}
\altaffiliation{These authors contributed equally}
\affiliation[Zaragosa]{Department of Analytical Chemistry, Universidad de Zaragoza, Spain}

\author{Samuel E. Rannala}
\affiliation[York]
{Department of Physics, University of York, United Kingdom}

\author{M. Puerto Morales}
\affiliation[Madrid]
{Materials Science Institute of Madrid (ICMM/CSIC), Spain}

\author{Sabino Veintemillas-Verdaguer}
\affiliation[Madrid]
{Materials Science Institute of Madrid (ICMM/CSIC), Spain}

\author{Carlton Jones}
\affiliation[Warrington]
{nanoTherics Ltd
Brookside Farm, Dig Lane, Warrington, WA2 0SH, United Kingdom}

\author{Lucía Gutiérrez}
\email{lu@unizar.es}
\affiliation[Zaragosa]{Department of Analytical Chemistry, Universidad de Zaragoza, Spain}

\author{Roy W Chantrell}
\affiliation[York]
{Department of Physics, University of York, United Kingdom}

\author{David Serantes}
\affiliation[Santiago]
{Applied Physics Department, Universidade de Santiago de Compostela, Spain}
\affiliation[Santiago2]
{Instituto de Materiais (iMATUS), Universidade de Santiago de Compostela, Spain}

\title{A device-independent approach to evaluate the heating performance during magnetic hyperthermia experiments: peak analysis and zigzag protocol}

\begin{document}

\begin{abstract}
Accurate knowledge of the heating performance of magnetic nanoparticles (MNPs) under AC fields is critical for the development of hyperthermia-mediated applications. Usually reported in terms of the \textit{specific loss power} (SLP) obtained from the temperature variation ($\Delta{T}$) \textit{vs.} time (t) curve, such estimate is subjected to a huge uncertainty. Thus, very different SLP values are reported for the same particles when measured on different equipment/laboratories. This lack of control clearly hampers the further development of MNP-mediated heat-triggered technologies. 
Here we report a device-independent approach to calculate the SLP value of a suspension of MNPs: the SLP is obtained from the analysis of the peak at the field on/off switch of the $\Delta{T}(time)$ curve. The measurement procedure, which itself constitutes a change of paradigm within the field, is based on fundamental physics considerations: specifically to guarantee the applicability of Newton's law of cooling, as i) it corresponds to the ideal scenario in which the temperature profiles of the system during heating and cooling are the same; and ii) it diminishes the role of coexistence of various heat dissipation channels. Such an approach is supported by theoretical and computational calculations to increase the reliability and reproducibility of SLP determination. This is experimentally confirmed, demonstrating a reduction in SLP variation across 3 different devices located in 3 different laboratories.
Furthermore, the application of this peak analysis method (PAM) to a rapid succession of field on/off switches that result in a \textit{zigzag}-like $\Delta{T}(t)$, which we term the zigzag protocol, allows evaluating possible variations of the SLP values with time or temperature.
\end{abstract}

\section{Introduction}
Magnetic nanoparticles (MNPs) as heat mediators have been proposed  for a wide range of heat-triggered medical applications such as cancer treatment \cite{Ortega2013,mahmoudi2018magnetic}, drug release\cite{guntnur2022demand}, gene activation,\cite{moros2019triggering} and controlled rewarming after cryopreservation protocol.\cite{chiu2021perfusion} Nanometer size iron oxide MNPs, based on their good biocompatibility, provide the ideal localised heat source inside any tissue or organ that can be activated remotely by AC magnetic fields\cite{soetaert2020cancer}. The efficiency of this process, referred to as magnetic hyperthermia (MH), is quantified in terms of the \textit{specific loss power} (SLP), i.e. the amount of electromagnetic energy that is converted into heat. However, determining the SLP value of a specific material with accuracy and low uncertainty is not free of difficulties.\cite{wells2021challenges, Wildeboer_2014} In addition to SLP, the heating properties of the MNPs are also sometimes described by a different parameter called the \textit{intrinsic loss power} (ILP). The term SAR (\textit{specific absorption rate}), used very extensively in the MH literature, should better be kept to describe power dissipation in tissues. \cite{wells2021challenges} 
Nevertheless, the alternative approach presented here can generally be applied for calorimetric measurements of such properties, referred to either as SAR or SLP.

The most extensively used approach to determine the SLP value of a magnetic nanoparticle suspension is the use  of calorimetric methods using the temperature variation ($\Delta$T) \textit{vs.} time (t) curve \cite{garaio2014wide}.  However, when comparing SLP values obtained by different laboratories problems start to arise, as the measurement of the heat released constitutes in itself a rather complicated task to reproduce. \cite{lemal2017measuring,soetaert2017experimental}  In fact, a recent study done in 21 different laboratories reported large variations between laboratories in the heating capability of a single batch of particles.\cite{wells2021challenges}  Such discrepancies mainly originate from differences in the measurement setups used in each laboratory. 

Most of the devices designed to measure the temperature variation $\Delta$T (t) curve when the MNPs are exposed to the AC field are non adiabatic.\cite{wells2021challenges} Therefore, the heat losses that appear during the measurement may be significantly different depending on the design of the device. Moreover, several works have already described that different mechanisms of heat losses can coincide within a given setup \cite{iglesias2021magnetic} and can have different timescales~\cite{coisson2016specific}. 
 
As a result from all the discussed problems, researchers working in the field of magnetic hyperthermia still lack a reliable and precise method to accurately determine the SLP value of a given particle suspension. Therefore, the development of an alternative approach, less dependent on the measurement devices and able to unify how the SLP values are calculated in a precise and reproducible way, becomes critical.

The objective of the current work is to present a  measurement protocol that diminishes the dependence on the specific device characteristics and environmental conditions. 
The work is presented in four sections. First, the theoretical framework behind the usual calorimetric methods is outlined, followed by a  review of the most widely used data analysis approaches to determine SLP. This first section also includes a comparison of experimental measurements and SLP data calculated from measurements performed in three different laboratories, as an illustrative example. 
The second section describes the origin of the problems that affect the SLP determination using current methods, focusing on the coexistence of various heat loss mechanisms and the inhomogeneous heating of the sample. We have performed a combined experimental/theoretical effort aimed at; i) differentiating effects attributable to the particles themselves from those defined by the device thermal properties, and ii) disentangling overlapping heat-loss effects on the determination of the heating performance, so that their roles may be minimized. 
The third section is devoted to the description of the new protocol to determine SLP. The proposed protocol, which we refer as the "zigzag protocol", is based on a set of repeated short time heating-cooling cycles and the subsequent analysis of the peaks arising when the AC field is switched off. We show that it is more beneficial to shift the SLP determination from the initial time of the heating curve, to the transition between heating and cooling. This is because the difference in losses during the heating and cooling phases are minimised close to the field on/off transition, and this allows to determine and subtract the correct heat loss contribution. Thereby, a more precise determination of SLP values is obtained. The final section provides validation of the proposed SLP calculation methodology. The theoretical validation includes the numerical generation of a test case where SLP values vary over time, to calculate the error in the SLP value. We then describe validation via an experimental inter-laboratory comparison in which three different devices are used to characterize the same magnetic nanoparticles, showing how the differences between SLP values estimated using the most common standard protocols, are diminished when the proposed \textit{zigzag} protocol is applied.

\section{The basics}

\subsection{Newton's law of cooling}
\label{sec:Newton}
In general, to have a correct description of the temperature evolution during the hyperthermia process, it is necessary to consider the detailed temperature profile both in time and space. This requires solving the heat diffusion equation including the sample (heat source region), the container and the surrounding environment:
\begin{align}
   \rho_r c_r  \frac{\partial \Delta T_r}{\partial t} &= k_r\frac{\partial^2 \Delta T_r}{\partial  r^2}+S,
   \label{eq:heat_eq3D}
\end{align}
where $\Delta T_r=T_r-T_{amb}$, provides the time evolution of the temperature ($T_r$), in relation to a surrounding medium ($T_{amb}$). $\rho$ is the density, $c$ the heat capacity, $k$ is the thermal conductivity  and $S$   is the heat source term. The subscript "r" indicates that Eq. \eqref{eq:heat_eq3D} needs to be considered in each point in space. 

The existing calorimetric methods used to determine the SLP value, which we will refer to as the "classical models", are based on simplified temperature dynamics, where the temperature evolution is defined in terms of Newton's law of cooling as \cite{LANDI201314}:
\begin{equation}\label{Eq.ODE}
\frac{dT}{dt} = -a(T-T_{amb})+S,
\end{equation} where $a$ is a phenomenological heat loss coefficient of the system, and $S$ defines the heating source ($\sim$ SLP) and $T_{amb}$  is the environmental temperature.

Note that Eq. \eqref{Eq.ODE}  requires the temperature within the heat source region to be uniform. In other words, under this assumption, the complex temperature profile inside and around the sample (Eq. \eqref{eq:heat_eq3D}) is replaced by just two temperatures: one for the  sample, $T$, and a the second one for the surrounding medium, $T_{amb}$.
Assuming that both the heat source ($S$) and the heat loss coefficient ($a$) are constant over time, the SLP value should be obtainable from the time evolution of the heating curve, which from Eq. \eqref{Eq.ODE} results in:
\begin{equation}
T(t)=T_{amb}+\frac{S}{a}\left[{1-e^{-at}}\right],
\label{Eq.T-heating}
\end{equation}
where:
\begin{equation}\label{Eq.S-parameter}
S=\frac{SLP\cdot{\rho_{NP}}}{\left({\frac{1}{c_{vol}}-1}\right)\rho_{w}c_{w}+\rho_{NP}c_{NP}}.
\end{equation} In Eq. \eqref{Eq.S-parameter}, $\rho$ and $c$ stand for the density and specific heat, respectively, of the nanoparticles (NP) and the dispersion medium ($w$ stands for water), and $c_{vol}$ corresponds to the volume fraction ($\%$) occupied by the particles.

\subsection{Usual methods to determine $SLP$: diversity of results}

Since $S=\lim_{t \to 0 }\frac{dT}{dt}$, and $S$ is uncorrelated from the thermal losses of the system (namely the "a" parameter), determining SLP from the initial slope of the $\Delta{T}(t)$ curve ($\Delta{T}(t)=T(t)-T_{amb}$) would allow, in principle, to obtain the intrinsic SLP value while neglecting the role of the thermal characteristics of the measurement device. Such an approach constitutes the so-called \textbf{initial slope method (ISM)} , where the SLP can be computed from the linear fit to the initial  $\Delta{T(t)}$ curve. In this approach the adiabaticity assumptions imply that, at the initial time, heat losses can be neglected. Such estimates are subject to huge uncertainty due to dynamic changes in the heat loss mechanisms inside the sample and even from the measurement device/environment. In other words, both the characteristics of the experimental set up and time frame in which the slope of the $\Delta{T(t)}$ curve is calculated will have a strong impact on the SLP estimation, leading to unreliable values also with very high uncertainties associated.

\begin{figure}[th!]
    \centering
    \includegraphics[width=1\linewidth]{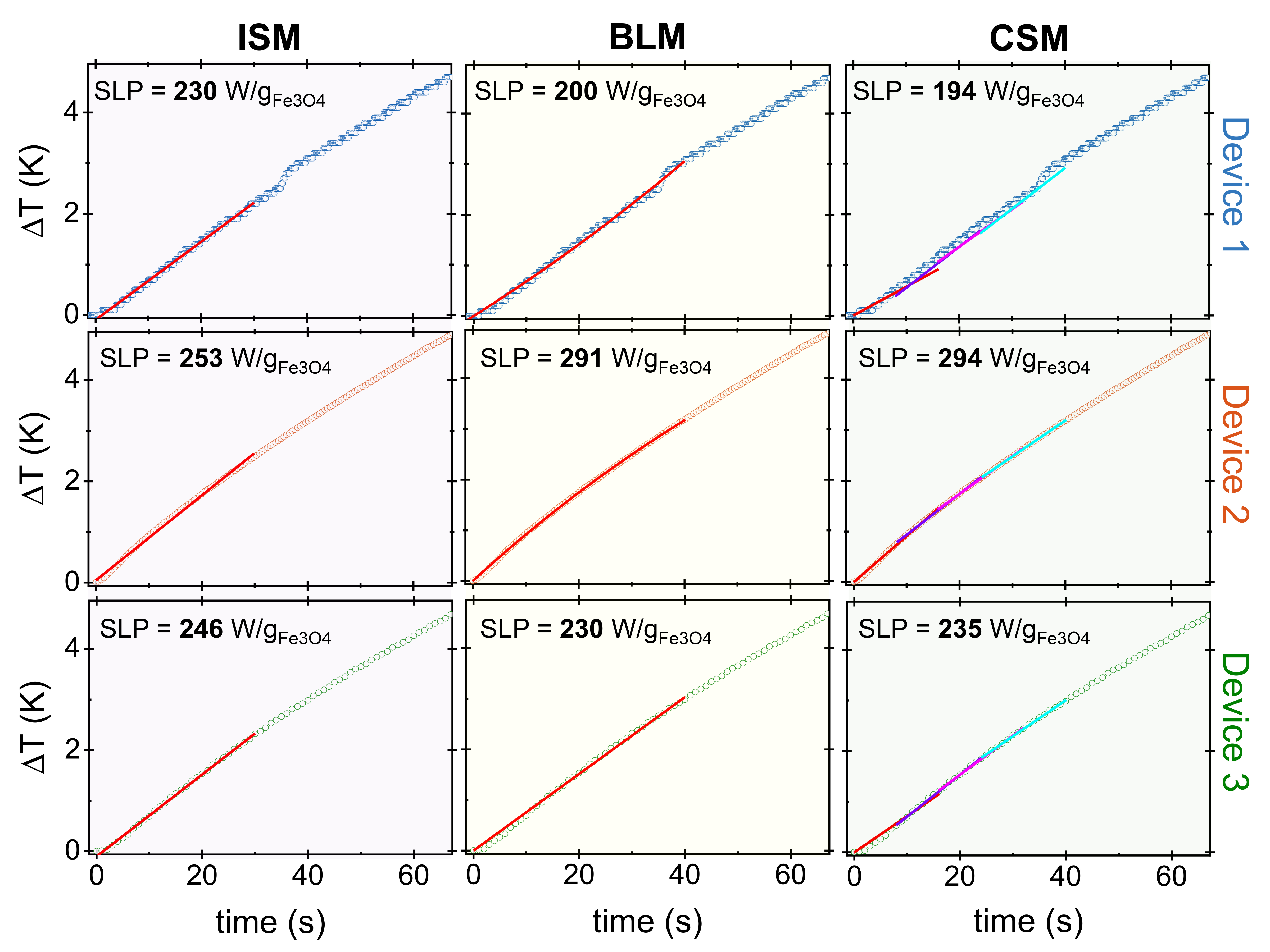}
    \caption{ Description of the different SLP-determination analysis, with three different measurement devices located in three different laboratories. Measurements were performed using a 1 mL suspension of the same particles (dextran coated, 32.0$\pm$6.7 nm of average size) at a concentration of $1mg_{Fe}/mL$. AC field conditions were 163.3 kHz and 35 mT for device 1, 165 kHz and 35 mT for device 2 and 172.4 kHz and 35 mT for device 3. The solid lines are the fitting curves. Details on the calculations can be found in Section 2 from Supporting information.}
    \label{fig:Classical}
\end{figure}

In order to solve some of the problems associated with the ISM, other alternatives based on a more complex analysis of the initial slope have been proposed. A detailed discussion on such methods can be found in Wildeboer \textit{et al.} \cite{Wildeboer_2014}. Briefly, the \textbf{Box-Lucas method (BLM)} uses a different equation to adjust the initial slope. \cite{Lanier2019} Alternatively, the \textbf{Corrected Slope Method (CSM)}  analyzes several time intervals of the initial slope to extract the SLP data. To illustrate how the use of these different data analysis affects the calculation of the SLP value, we have analyzed a suspension of magnetic nanoparticles (dextran coated, 32.0$\pm$6.7 nm of average size, see Section 1 from the Supporting information for further characterization details) using the ISM, BLM and CSM approaches; see Figure \ref{fig:Classical}. In addition, the samples were measured in three different devices operating in  very similar conditions (163.3 kHz and 35 mT for the device 1, 165 kHz and 35 mT for the device 2, and 172.4 kHz and 35 mT for device 3, in all cases using a suspension of 1 ml with an iron concentration of 1 mg/mL). For a given device, significant differences, up to 16\%, were observed in the SLP values calculated by the different methods (see specific values in Figure \ref{fig:Classical}). When comparing devices, SLP values calculated from device 2 were  significantly larger than those obtained with the other two devices, the lowest obtained value (194 W/g Fe$_3$O$_4$) being for the CSM with Device 1 and the highest for the same method in Device 2 (294 W/g Fe$_3$O$_4$), which corresponds to a 34\% difference on the obtained values. From this analysis, it is clear that there is a large uncertainty in the SLP value associated with both the device used for the measurements and also the method selected for the data analysis.

In addition to the data analysis methods focusing on the analysis of the initial slope of the heating curve, an alternative approach is the \textbf{Decay Method (DM)}, that includes in the sample characterization recording both the heating curve, when the AC field is switched ON, and the cooling curve, once the AC field is switched OFF. 
Observing the different curves displayed in Figure \ref{fig:Classical}, where small changes (even abrupt jumps) in the data may lead to significant changes in slope, one may easily imagine that extending the fitting range would help minimising the dependence on the specific features of the curves. Furthermore, given that there are only two unknown values in Eq. \eqref{eq:heat_eq3D}, it might seem reasonable to devise two different scenarios to fit two similar curves, so that we have two equations with two unknowns.
Therefore, the \textbf{Decay Method} uses the the cooling phase to obtain a characteristic time of the system cooling down and the steady state temperature. However, from an experimental point of view, this method requires longer characterization times, to assure that the steady state temperature has been reached and also to characterize the cooling phase, therefore, this approach is much less frequently used than the those described earlier. However, it is not just a matter of time: as will be explained in the next section, the fit of the cooling curve is not straightforward and, thus, the use of the Decay Method to obtain the SLP value also has associated systematic errors.

\section{Identifying the problems - applicability of Newton's law?}\label{sect.IdentificationPROBLEMS}

\begin{figure}[bt!]
    \centering \includegraphics[width=0.7\linewidth]{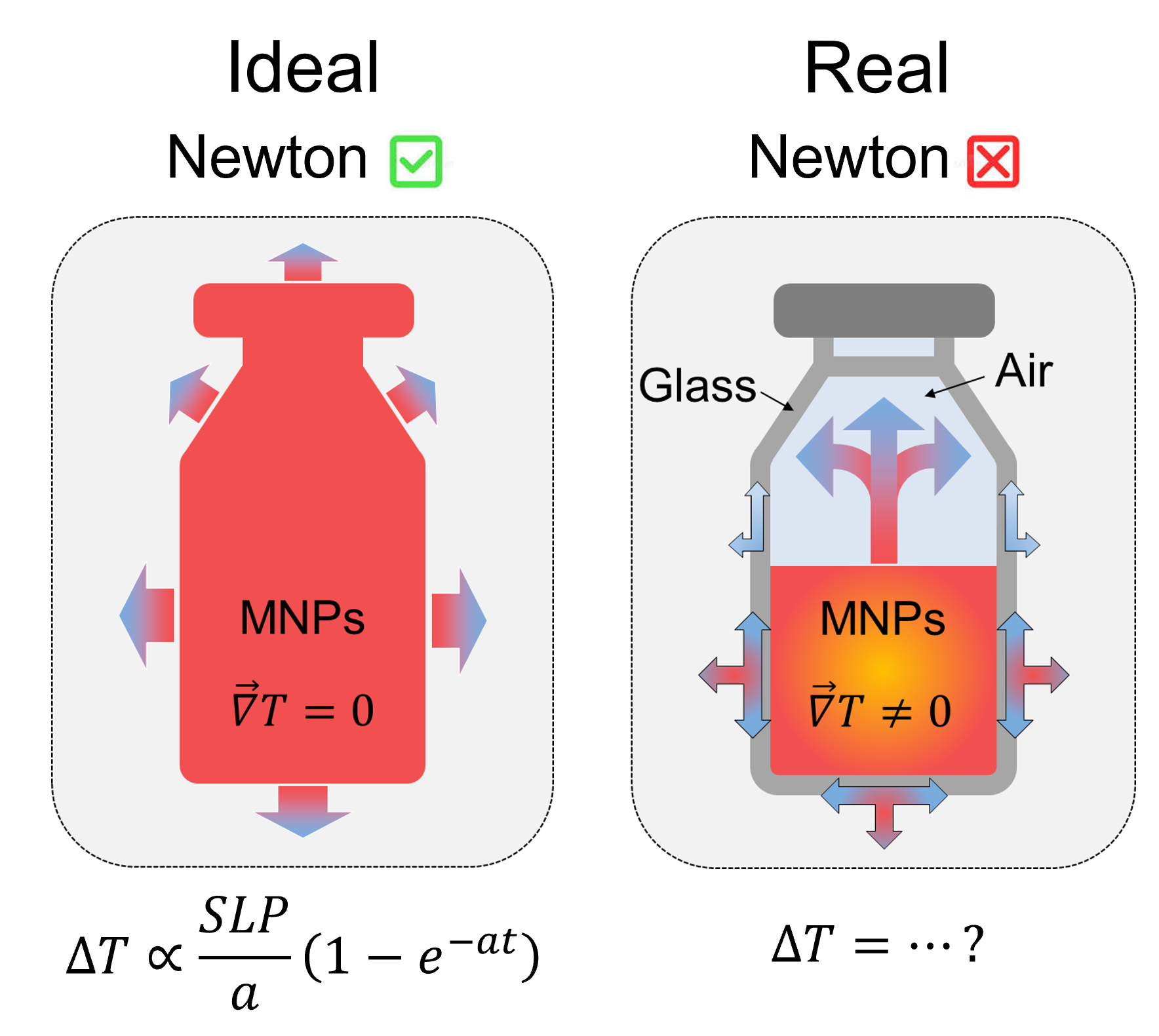}
    \caption{Schematic illustration of the differences between the ideal scenario under which Newton's law of cooling would be applicable (i.e. homogeneous temperature, and one heat-exchange mechanism; left panel); and the real situation in which the sample temperature is not homogeneous, and heat can be exchanged through different channels (right panel).}
    \label{fig:ideal vs real}
\end{figure}

Aiming to elucidate which might be the underlying reasons of the diversity in SLP values, not only between different devices but also when applying different protocols to the same data $\Delta T(t)$, in this section we  perform a detailed analysis of the underlying physical background: the assumption that Newton's law of cooling is applicable. Thus, we analyse in detail the two implicit simplifications assumed in deriving Eq. \eqref{Eq.T-heating} from Eq. \eqref{Eq.ODE}. First, that there is only one heat-loss channel, defined by the "a" parameter. Second, that the temperature of the sample is homogeneous. The differences between the ideal situation assumed in Newton's law and the real system are schematically depicted in Figure \ref{fig:ideal vs real}

\subsection{Assumption 1: a single heat-loss channel}\label{subsect.one_loss_channel}

\begin{figure}[bt!]
    \centering \includegraphics[width=0.7\linewidth]{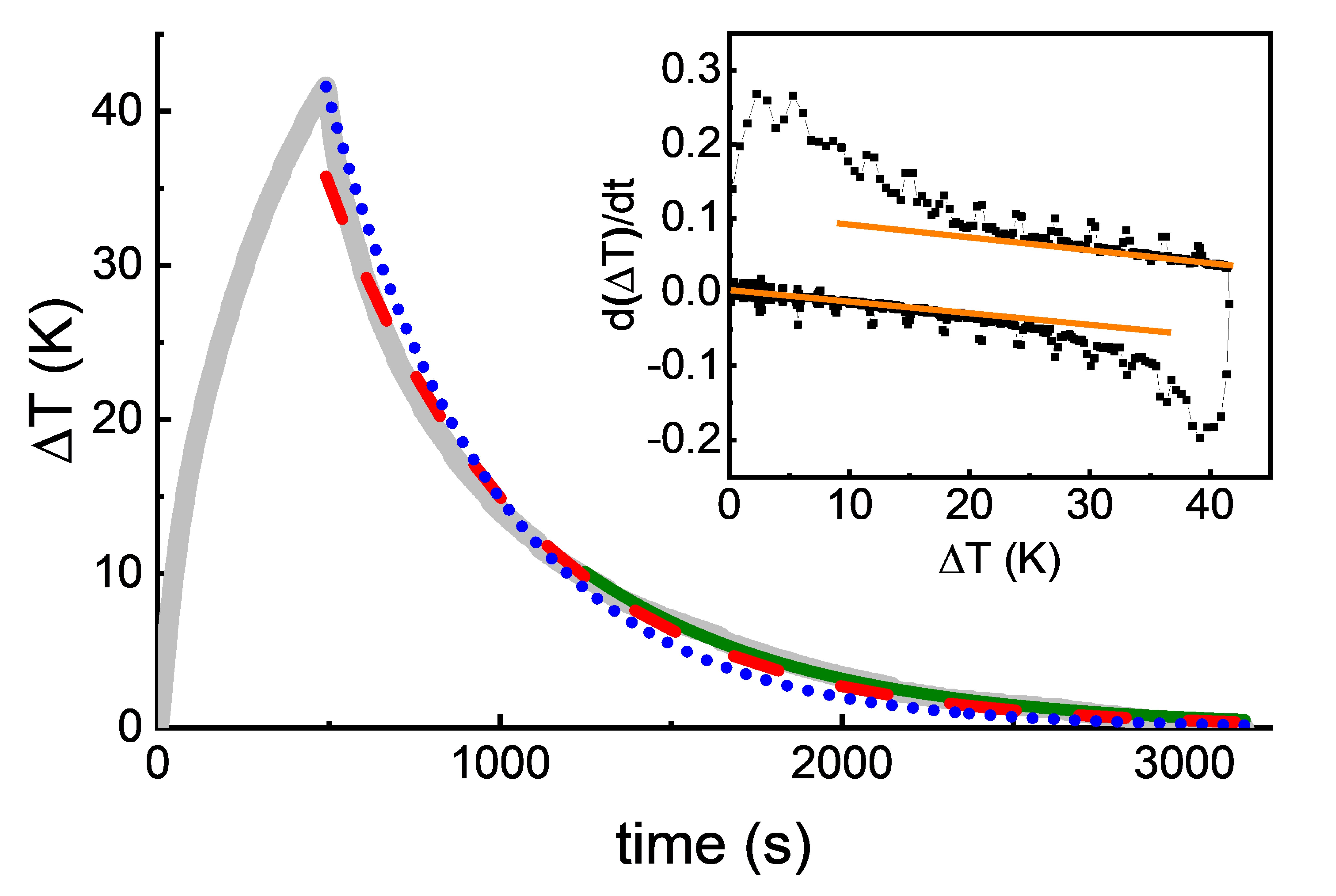}
    \caption{Single heating and cooling cycle measurement of the same particles from Figure\ref{fig:Classical}. Three different approaches to fit the cooling part using a single exponential have been performed. The blue fit (dotted line) was performed using the initial time part of the cooling curve, obtaining $a=0.0020$. The red fit (dashed line) was a free fit to the entire cooling part, with $a=0.0015$. The green fit (continuous line) was obtained from the final part of the curve, with $a=0.0017$. (inset) Representation of the $\frac{d(\Delta{T})}{dt}$ \textit{vs.} $\Delta{T}$ of the same data. If a single heat loss mechanism was occurring, the data should be a straight line according to Eq. \eqref{Eq.ODE}. Results indicate that there is not a single heat loss mechanism. Straight lines are a guide for the eye. Similar curves were observed in ref.\cite{Hanson2023}. }
    \label{fig:cooling}
\end{figure}

The first simplification performed to reach Eq. \eqref{Eq.T-heating} from Eq. \eqref{Eq.ODE}, assumes that there is only one heat-loss channel, defined by the "a" parameter. As explained in the previous section, the \textbf{Decay Method (DM)} uses the cooling phase (where SLP = 0, as no AC field is applied) to obtain the phenomenological heat loss coefficient  "a", and take that value to obtain the SAR from the heating phase. Since Newton's law of cooling, Eq. \eqref{Eq.ODE}, corresponds to a single exponential decay, the exponential fit of the cooling part would provide the "a" value that could be used to fit the heating part. However, as discussed in the insightful work by Landi \cite{LANDI201314}, it is often observed that a single exponential fit does not match the experimental cooling part. This is illustrated in Figure \ref{fig:cooling}, where a single "a" value to fit the cooling part of the $\Delta{T(t)}$ cannot be obtained in a measurement performed using the same particles described in the previous section. In fact, depending on how the fit of the cooling part is performed, very different "a" values are obtained (see the caption of Figure \ref{fig:cooling} and the Supporting Information file for details). The lack of a single "a" value, indicative of the presence of several heat-loss mechanisms, is more clearly emphasized if plotting $\frac{d(\Delta{T})}{dt}$ \textit{vs.} $\Delta{T}$, which according to Eq. \eqref{Eq.ODE} should be a straight line if a single heat loss mechanism is occurring. As the inset in Figure \ref{fig:cooling} shows, this is clearly not the case, as recently analysed in detail by Hanson \textit{et al}. \cite{Hanson2023}. The heat loss mechanisms and the corresponding "a" parameter are highly influenced by the device setup and environmental conditions, which can differ significantly from one laboratory to another. The presence of several heat-loss mechanisms has also been recently reported by Iglesias \textit{et al.} \cite{iglesias2021magnetic}
In summary, several heat loss mechanisms may simultaneously occur during the calorimetric methods to determine the SLP value, being a source of the variable results reported in the literature.

\subsection{Assumption 2: homogeneous temperature}\label{subsect.homogeneous_temperature}

\begin{figure}[bth!]
\centering
\includegraphics[width = 0.5\columnwidth]{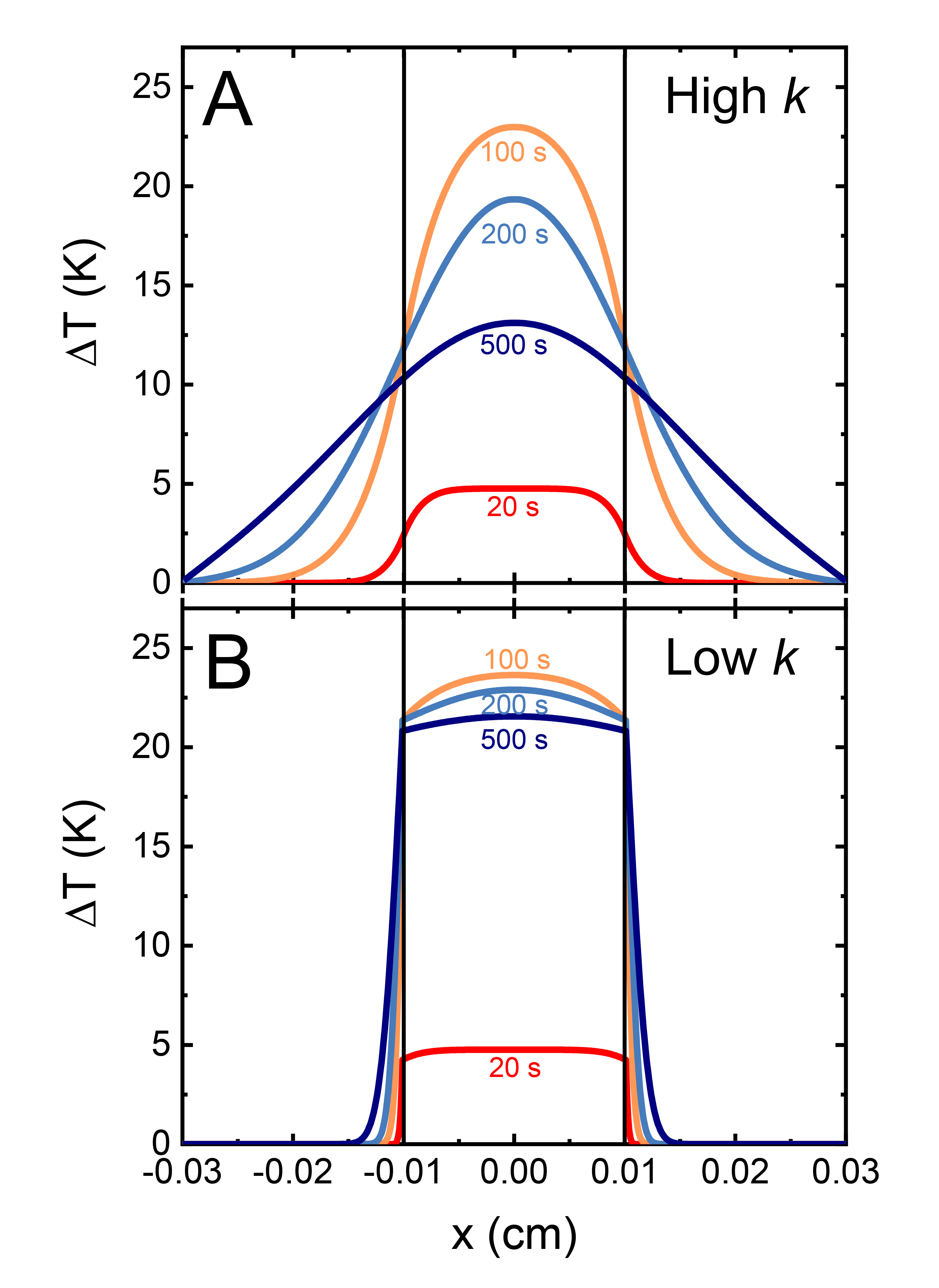}
\caption{Plots of the temperature profile during heating and cooling. Vertical lines show the extent of the heated fluid.  (A) High thermal conductivity of the vessel ($k_{wall}=0.8$ $Wm^{-1}K^{-1}$)  and (B) low thermal conductivity ($k_{wall}=0.01$ $Wm^{-1}K^{-1}$) of the vial. The effect of improved insulation as expected leads to a larger temperature rise, and additionally a lower temperature gradient within the vial. }
\label{fig:heating-cooling_sketches}
\end{figure}

The second simplification performed to reach Eq. \eqref{Eq.T-heating} from Eq. \eqref{Eq.ODE} is that the temperature of the sample is homogeneous. Essentially this requires the temperature within the vial to be uniform, which has been observed experimentally to not be the case, as temperature gradients are clearly observed when using infrared cameras to monitor the whole volume of the sample \cite{valdes2023thermographical,espinosa2018magnetic}.
We investigate this non-uniform heating within the sample by simulating the heating and cooling processes using a simple 1D heating model (Figure \ref{fig:heating-cooling_sketches}). The reality of heat loss processes in practical experimental set-ups is complex, possibly involving more than one heat-loss process,  and difficult to model. Our approach was intended to illustrate the heating and cooling processes using a simple and physically transparent model, which is successful in highlighting the need for an advanced measurement protocol and evaluating its likely efficacy prior to experimental validation. We modeled the increase/decrease of temperature with the SLP driving term on/off. The conjecture was as follows: when the AC field is ON, the SLP is localised giving rise to a rapid (uniform) heating with losses mainly through the boundary of the sample holder, whereas, when the field is OFF, during the cooling phase, there will be a slow migration of heat out of the system due to small gradients within the sample holder. In 1D the time variation of the temperature (with $\Delta T=T-T_{amb}$) is given by:
\begin{align}
   \frac{\partial \Delta T}{\partial t}&= \alpha \frac{\partial^2 \Delta T}{\partial  x^2}+{S},
   \label{eqn:heat_eq_1D}
\end{align} 
where  $\alpha = k/(\rho C)$ is the diffusivity, and for simplicity of notation will use $S=\frac{SLP}{\rho c}$. 
Eq.~\eqref{eqn:heat_eq_1D} is solved numerically with interface conditions between the fluid and the vial corresponding to heat loss by conduction and by convection. These  conditions are given in the methods section and in Section 3 of the Supporting Information.

Results from this analysis can be found in Figure \ref{fig:heating-cooling_sketches}, which shows plots of the temperature profile during heating and cooling. Vertical lines show the extent of the heated fluid.  Data are shown for high thermal conductivity of the vessel ($k_{wall}=0.8$ $Wm^{-1}K^{-1}$)  and low thermal conductivity ($k_{wall}=0.01$ $Wm^{-1}K^{-1}$) of the vial.
As expected, it can be observed that the temperature gradients occurring within the liquid sample differ significantly between the heating and cooling phases as a consequence of the heat flow and heat loss processes. Moreover,  the impact of the thermal conductivity values of the vial has also been tested, see Figure \ref{fig:heating-cooling_sketches}. In the case of high thermal conductivity of the vial the temperature gradient varies strongly across the heated fluid and also varies considerably with time. As expected the increased insulation leads to a higher temperature rise. It also leads to profiles with rather less curvature. 

These results illustrate the fact that heat losses  and temperature gradients within the sample can be strongly dependent on the details of the experiment and in particular the thermal properties of the container. 
This clearly indicates that validity of Newton’s law of cooling cannot be generally assumed to be true.

\section{A new approach to calculate the SLP}

In order to solve the problems associated to the presence of temperature gradients and the co-existence of simultaneous heat loss mechanisms described in the previous section, we  have generated an alternative protocol to perform the calorimetric experiments and the data analysis using which it is possible to significantly diminish the role of the heat-losses mechanisms associated with the different devices. This protocol is based on the following principles. First, there is a shift in the data that is going to be analyzed, from the the initial slope measured in the classical methods, to the peak generated when the field is switched off. Second, instead of doing a single measurement, a series of on/off switches of the field are performed resulting in a zigzag shape $\Delta{T}(t)$ curve. The reasons behind these two modifications are explained in detail below.

\subsection{Change of paradigm: the peak analysis}\label{subsect.peak}

\begin{figure}[bth!]
\centering
\includegraphics[width = 0.9\columnwidth]{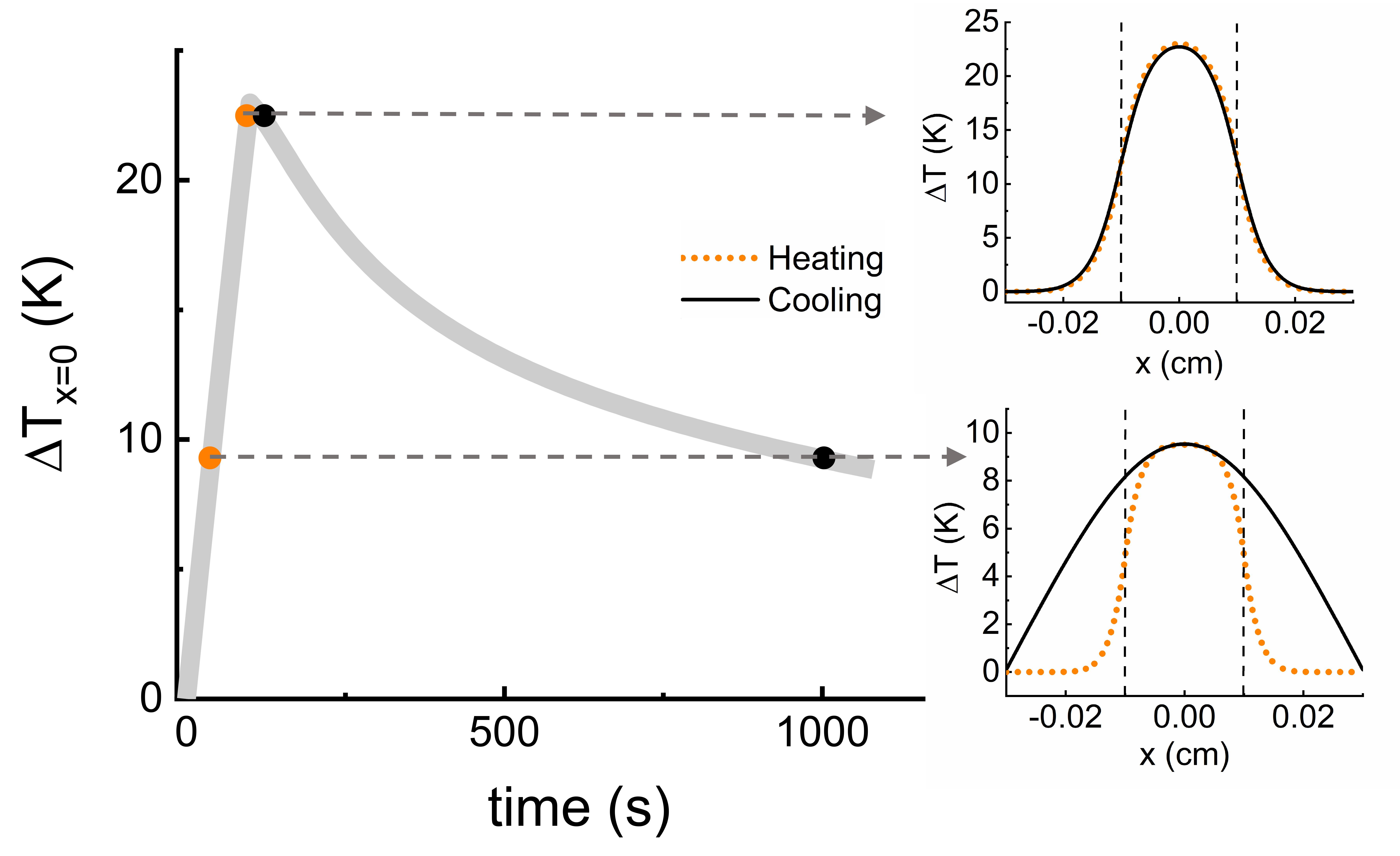}
\caption{Plots of the temperature profile during a single heating and cooling experiment. The vertical lines show the extent of the heated fluid. Near the peak, the gradient in the heating curve and the cooling curve is similar. Further away from the peak, when the temperature of the suspension is 9.5 ºC, larger differences in the temperature gradients inside the vessel are observed.}
\label{fig:heating-cooling gradients}
\end{figure}

The main reason to shift the data analysis from the initial slope to the peak generated by the AC field on/off switch is that if we were able to measure the cooling curve immediately after heating, the effect of losses at the transition between heating and cooling phases and also the coexistence of longer relaxation-losses mechanisms on SLP determination would be minimised. Further, as we show later, by measuring the decay curve it becomes possible to correct accurately for the heat losses thereby removing a significant systematic error in SLP measurements.  This is illustrated in Figure \ref{fig:heating-cooling gradients}, where the 1D model described in the previous section has been used to plot the gradient along the sample at some specific time points during the heating and the cooling parts along the $\Delta T(t)$ measurement. We show that, when we are looking at data far away from the peak,  the temperature profiles are very different at a given temperature during the heating and cooling phases  In contrast, when we are closer to the peak, these gradients are similar and the form of the temperature profile is essentially unchanged around the peak temperature.

In fact, if we write the heat diffusion equation for the two process, 1) heating when the AC field is on:
\begin{align}
    \left| \frac{\partial \Delta T_r}{\partial t} \right |_{heating} &= \alpha_r  \left|\frac{\partial^2 \Delta T_r}{\partial  r^2}  \right |_{heating}+S
   \label{eq:heat_eq3D_AF_on} ,
\end{align}
 and 2) cooling, when AC field is off:
\begin{align}
      \left| \frac{\partial \Delta T_r}{\partial t}  \right |_{cooling}&= \alpha_r  \left| \frac{\partial^2 \Delta T_r}{\partial  r^2} \right |_{cooling}
   \label{eq:heat_eq3D_AF_off} .
\end{align}
A reasonable assumption, validated by the simulations, is that at the transition between heating and cooling the spatial derivatives will be  very similar, thus subtracting  Eq. \eqref{eq:heat_eq3D_AF_on} from Eq. \eqref{eq:heat_eq3D_AF_off} leads to as precise as possible  determination of S and implicitly SLP:
\begin{align}
    S=\left| \frac{\partial \Delta T_r}{\partial t} \right |_{heating} - \left| \frac{\partial \Delta T_r}{\partial t}  \right |_{cooling} .
   \label{eq:zigzag_protocol}
\end{align}
Therefore, shifting the data analysis from the initial slope to the peak seems a feasible way of avoiding the problems associated with the temperature gradients that originated within the sample during the heating and cooling processes.

\subsection{Repeated AC field ON/OFF switches}\label{subsect.ac switches}

The peak analysis proposed in the previous section could be done for one heating-cooling cycle or multiple heating-cooling cycles.  Other groups have already proposed the use of a "stepped heating procedure" \cite{iacob2015stepped} to acquire heating and cooling data at different temperatures or to verify the absence of parasitic signals from the thermocouple \cite{liu2022field}. In addition, the idea of applying the AC magnetic field intermittently has been also recently described as a way to control the global temperature reached in tissue phantoms \cite{tsiapla2021mitigation,carlton2023new}.

Our proposal is that, given that the data needed for the peak analysis described in the previous section corresponds only to the values immediately close to the peak, the temperature profile remains essentially constant and the SLP value can be determined using Eq.~\eqref{eq:zigzag_protocol}. It follows that there is no need to wait for the temperature to reach an equilibrium (neither in the heating, nor the cooling part), and a sequence of fast cycles of ON/OFF field can be performed. This approach will provide two main advantages. First, the calculation of SLP values in several "peaks", will allow calculating the error associated with the SLP determination in repeated measurements faster than repeating "classical methods" several times. Furthermore, it will allow tracking any possible changes in SLP values due to differences in the global temperature, also providing a tool to calculate the SLP at body temperature.

\begin{figure}[th!]
    \centering
    \includegraphics[width=0.5
    \linewidth]{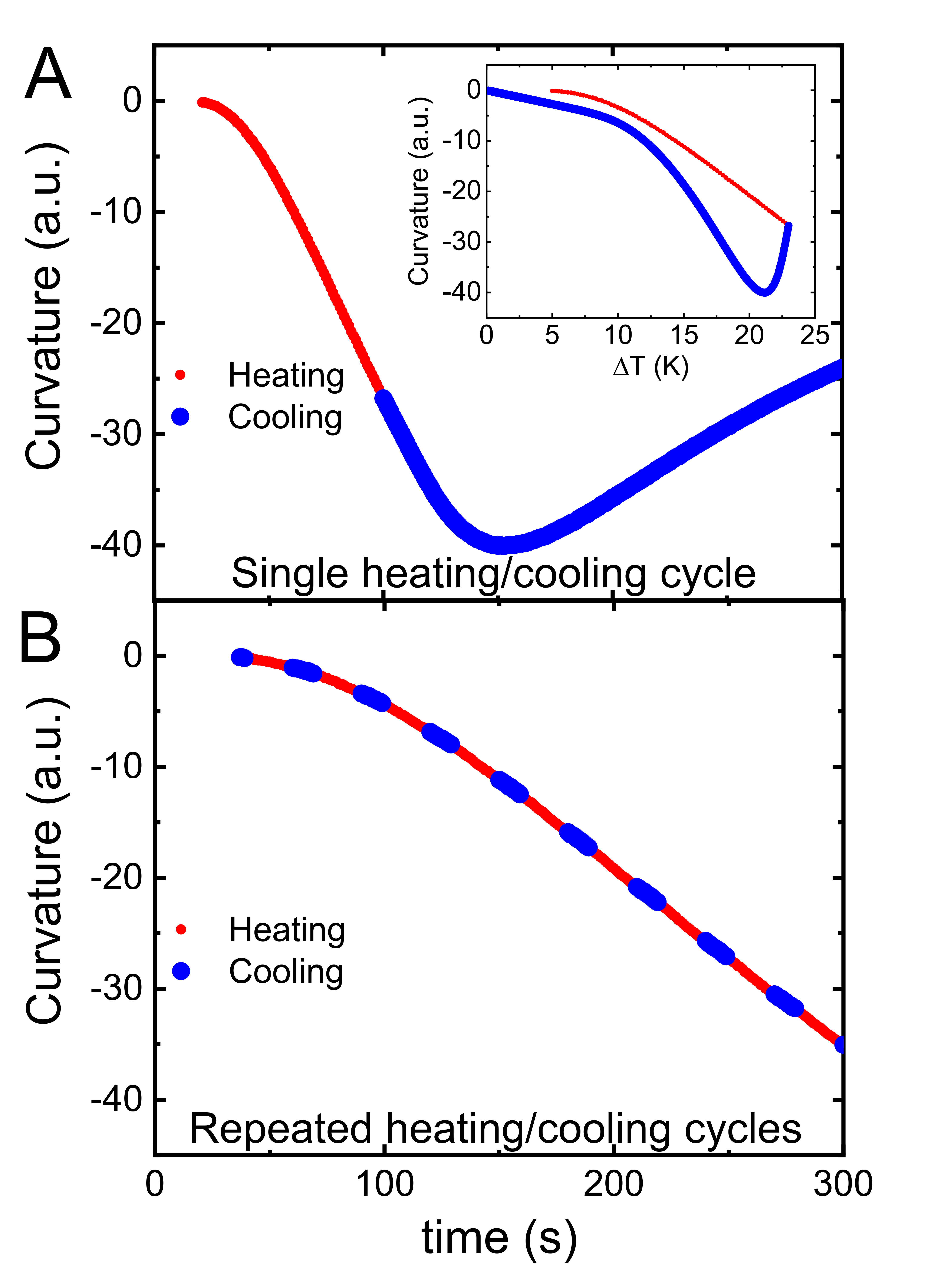}
    \caption{ Comparison between single heating/cooling cycle (A) vs repeated heating-cooling cycles (B). The plots show the curvature of the temperature profile inside the heated fluid. The negative sign indicates the concave profile of temperature and the value indicates the degree of homogeneity of temperature profile. The inset in (a) is the curvature as a function of temperature for AC field on (red: heating) and AC field off (blue: cooling). This shows a complex temperature evolution inside the fluid both during the heating and cooling process, which is not symmetric.}
    \label{fig:curvature}
\end{figure}

To verify the use of this pulsed AC field  approach, we have quantified the degree of temperature homogeneity inside the sample (as illustrated in Figure \ref{fig:heating-cooling_sketches}). The curvature of the temperature profiles as function of temperature is presented in (Figure \ref{fig:curvature}). We calculate a radius of curvature as given in the methods section. The characteristic radius of curvature is calculated at the center of the heated vial.

We have compared the temperature profile when performing a single heating and cooling cycle and the \textbf{zigzag protocol} using the simple 1D model outlined earlier. Because the radius of curvature is infinite at time $t=0$ we choose to characterise the curvature as the inverse of the radius of curvature calculated as given in the materials and methods section, Eq.~\eqref{eq:radius}. In the single heating/cooling cycle, it can be observed that there is a significant variation of the curvature during the heating process (red line in Figure \ref{fig:curvature}). Additionally a more complex variation of the curvature with time occurs when the field is turned off (the cooling phase). We can observe that the behaviour is not symmetric for heating and cooling, supporting our hypothesis that the heat-loss mechanisms are complex and can not be assumed to be the same during the two phases. As expected the curvature converges at the transition between heating and cooling, thereby supporting  the use of the peak analysis. Indeed, a more linear behaviour of the curvature is observed during the simulation of the zigzag protocol, where the convergence between the heating and cooling parts of the curve at the peak is clear.  These results support the idea of using the peak analysis when applying pulsed AC fields to the sample.

\subsection{The zigzag protocol}\label{subsect.zigzag}

\begin{figure}[bt!]
    \centering
    \includegraphics[width=1.0\linewidth]{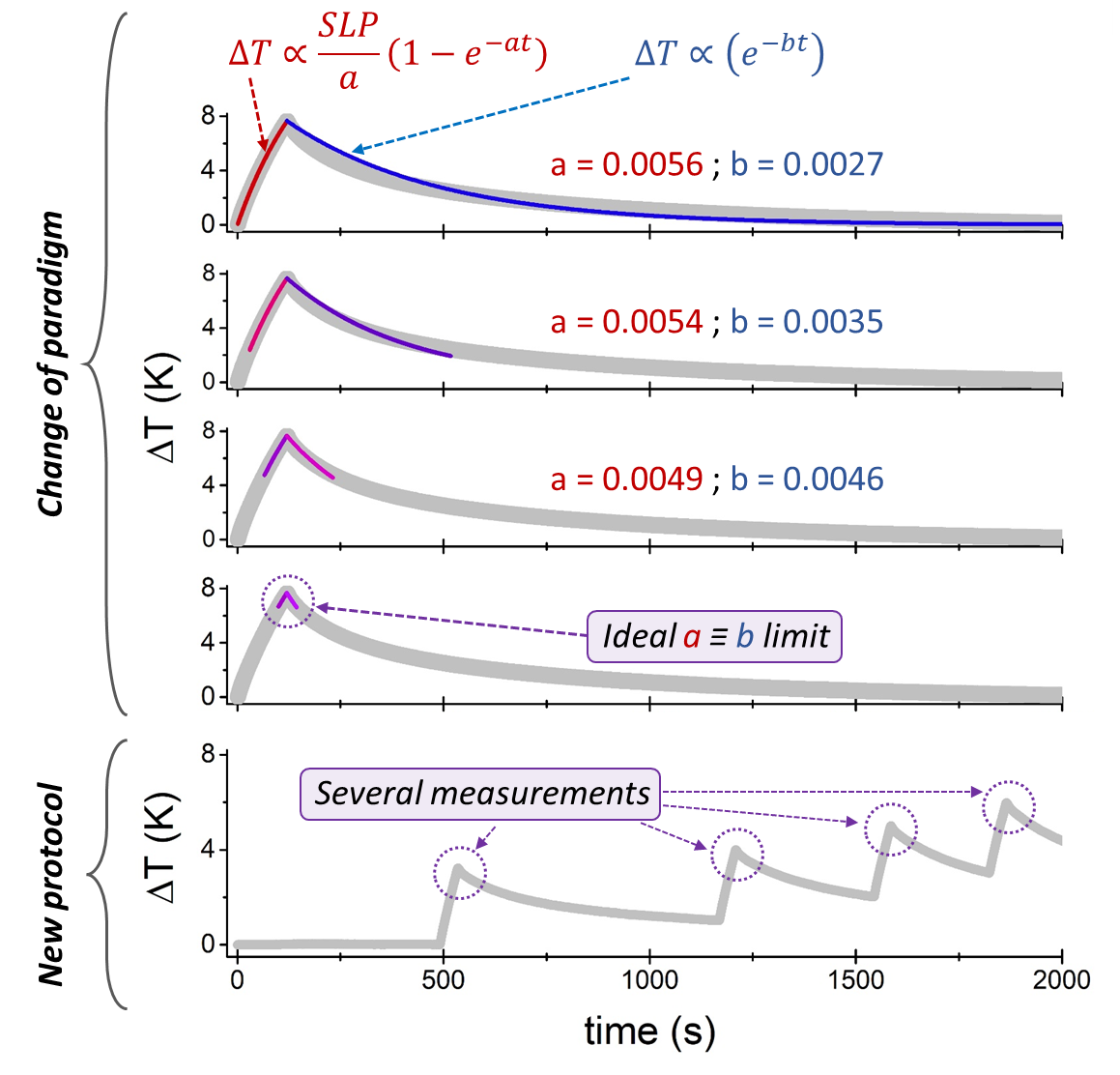}
    \caption{ Scheme depicting the advantages of the use of the Peak Analysis Method and the zigzag protocol. Shifting the data analysis to the peak resulting from the on/off magnetic field switch allows assuming similar heat loss mechanisms during the heating and the cooling phases. Repeating the on/off switches several times in a zigzag way allows quickly obtaining replicates and tracking the evolution of the heating capacity of the particles over time and temperature.} 
    \label{fig:change_of_approach}
\end{figure}

We have termed our proposal for the new approach to determine the SLP value the \textbf{zigzag protocol} (Figure \ref{fig:change_of_approach}). This proposal comprises two complementary ideas. First, shifting the data analysis from the initial slope to the peak. As described above, the peak analysis will allow studying the ideal limit where the heat loss mechanisms in the heating and cooling phases of the $\Delta{T}(t)$ curve approach  a single value, providing a scenario that allows the determination of a SLP value independent of the device. Moreover, performing a zigzag type of measurement, where the AC magnetic field is intermittently switched on and off,  it is possible to determine the error associated with the SLP value using different fast repetitions of the peak analysis,assuming a constant value of the SLP. A Standard Operating Procedure to perform this analysis has been developed and is provided in Section 4 of the Supporting Information.

\subsection{Time dependent SLP}

 \begin{figure}[bth!]
\centering
\includegraphics[width=0.9\linewidth]{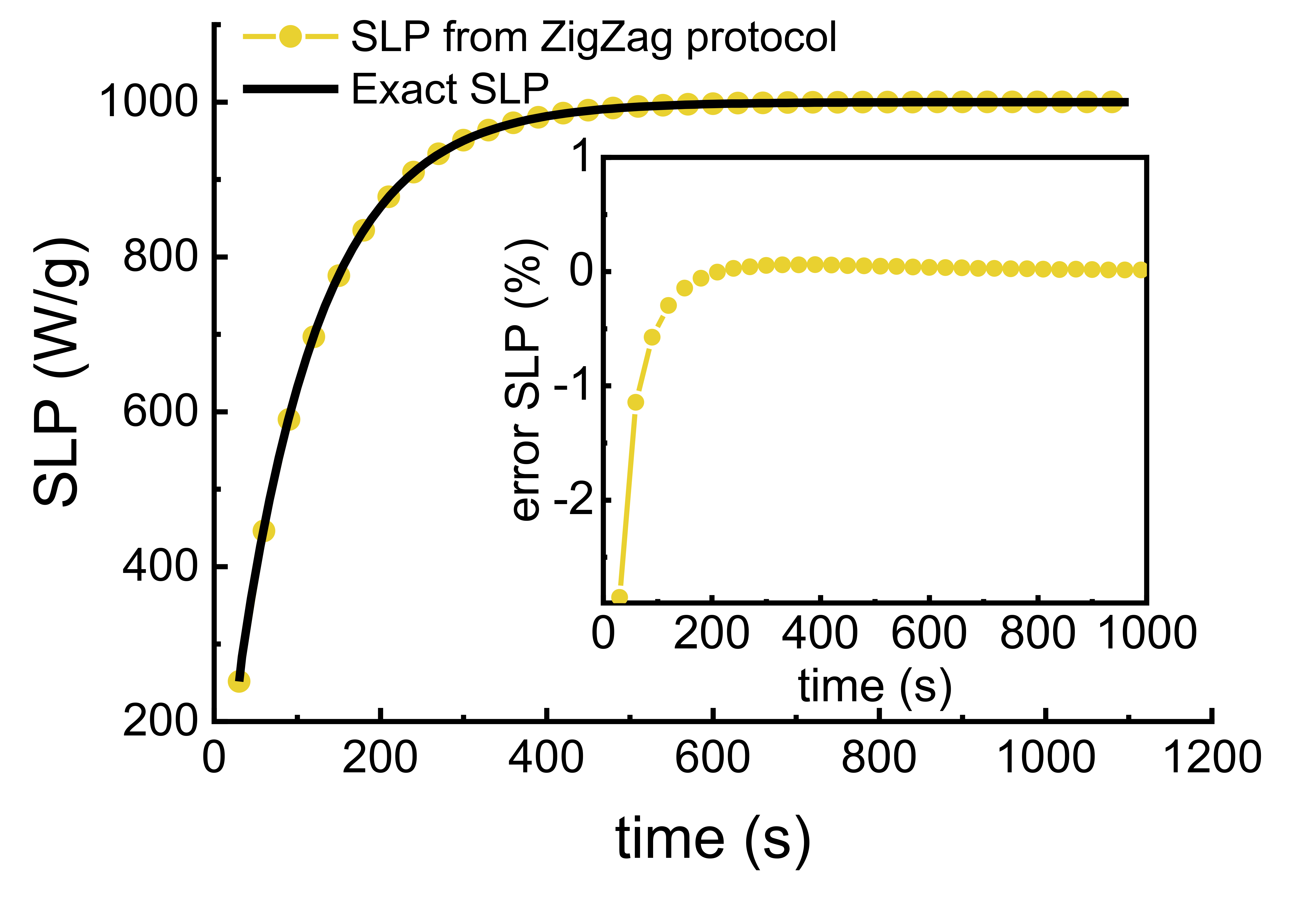} 
\caption{Example of time dependent SLP (black line, input SLP) and the values extracted from zig-zag method (gold dots). The results based on the error given in the inset, indicate a very accurate determination of the input SLP.  }
\label{fig:Sim_expSAR}
\end{figure}

We have shown that the PAM method provides determination of the SLP without systematic errors introduced by thermal transport processes.  In addition to reducing the errors on the SLP calculation, the use of the zigzag protocol has other interesting advantages. In the first instance, under the assumption of a constant (time invariant) SLP,  the extra data provided by the zigzag method allows further averaging and a reduction in the statistical error. 
Further, it could also be used to accurately determine a time varying SLP. This might arise from temperature variations in the intrinsic nanoparticle properties, for example the use of low Curie temperature materials to control their heat output which would give rise to a time (temperature) varying SLP \cite{brezovich1984temperature, Natividad2011}. A further example is the possibility of time dependent chaining as observed by Mille et.al\cite{Mille2021}. To illustrate this aspect, we simulated a case when SLP increases exponentially in the form $SLP=SLP_0 [1-exp(-t/t_0)]$. The results for $SLP_0=1000W/g$ and $t_0=100s$ are shown in Figure \ref{fig:Sim_expSAR}, indicating the capability of the zigzag method to accurately determine SLP, not just in the case of constant SLP but also for the case where SLP varies during the MH measurement.  We believe, this feature provides significant advantages of this protocol. In this way, any changes in the SLP due to particle clustering (e.g. chaining when AC field is on), or temperature dependence of magnetic properties could be tracked over time. However, it should be noted that this will require high accuracy (low noise) measurements which should be possible by using high resolution temperature probes and improved averaging. 

In summary, the zigzag protocol not only provides a better estimation of SLP, but can also 1) provide the variation of SLP during the heating protocol and/or 2) the SLP value at the desired operation temperature, which is generally different from the ambient temperature at which the "classical methods" based on the initial slope are generally applied.

\subsection{Experimental validation of the "zigzag protocol"}
\begin{figure}[bth!]
    \centering \includegraphics[width=1\linewidth]{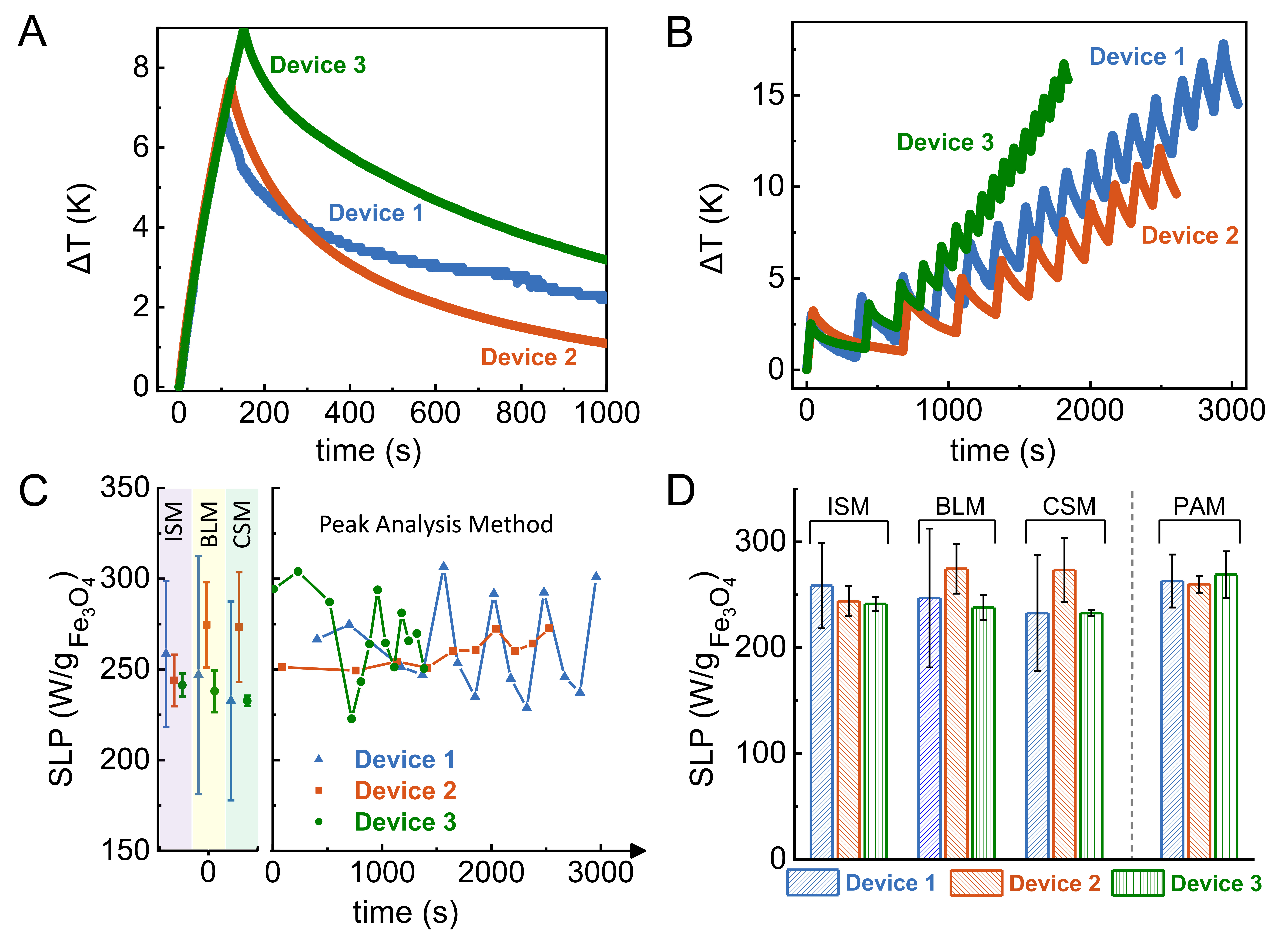}
    \caption{ SLP calculation from experimental calorimetry measurements using three different devices. (A) Classical single heating/cooling measurement (B) Repeated heating-cooling cycles (zigzag measurements). (C) SLP values calculated from the experimental results. ISM, CSM, and BLM approaches have been used to calculate the SLP values from several replicates of single heating/cooling measurements as the ones shown in panel (A). In contrast, the the Peak Analysis Method (PAM) has been applied to each of the peaks from the zigzag measurements shown in panel (B). (D) Comparison between the extracted SLP values for the three devices when using different methods for the data analysis. Note that SLP values for the PAM method correspond to the average value for all the peaks analyzed independently and shown in panel (C).} 
    \label{fig:SAR ZigZag}
\end{figure}

While the theoretical calculations provide insight and an essential verification, the final step is to proceed to experimental validation of the protocol  using the same particles described in the first section (see also Figure S1 from the Supporting information).
Suspensions of the dextran coated particles are characterized using three different commercial devices for the SLP analysis (Figure \ref{fig:SAR ZigZag}a and b). In all devices, a suspension of 1 ml with an iron concentration of 1mg/mL prepared from the same batch is used. In all devices, very similar AC field conditions are used (163.3 kHz and 35 mT for device 1 (D5 Series from nB nanoScale Biomagnetics), 165 kHz and 35 mT for device 2 (Fives Celes, MP 6kW) and 172.4 kHz and 35 mT for device 3 (Nanotherics)). The sample is measured in each device using two experimental protocols, a single heating/cooling cycle and the zigzag protocol, where several cycles of faster heating/cooling cycles are sequentially performed following the Standard Operating Procedure provided in the Supporting Information.

When considering the single heating-cooling cycle, measurements performed with the three devices revealed a fairly similar initial slope but differences in the cooling phases associated with different degrees of insulation and therefore thermal losses are evident (Figure \ref{fig:SAR ZigZag}a). Devices 2 and 3 have a more uniform cooling process, whereas device 1 has clearly at least two timescales associated with the cooling. As described in the first section, significant differences in the SLP values, up to 35\%, are observed in the calculations performed using the different classical methods applied to the initial slope analysis (see Figure \ref{fig:Classical}). Several repetitions of the measurements were performed in order to calculate the uncertainty associated with the SLP value calculation depicted in Figure \ref{fig:SAR ZigZag}c. This analysis clearly shows the large uncertainty in the SLP value associated with both the device used for the measurements and also the method selected for the data analysis.

Then, the zigzag protocol is applied to measurements with the three different devices (Figure \ref{fig:SAR ZigZag} b). We note that device 3 shows a much more rapid temperature increase than devices 1 and 2, suggesting a smaller heat loss. Figure \ref{fig:SAR ZigZag} c, shows the comparison of the SLP values obtained using the classical data analysis approaches (ISM, BLM and CSM) applied to the initial slope of the single heating/cooling cycle compared to the SLP values obtained using the peak analysis method for each of the sequential peaks generated in the zigzag protocol over time. Although not used in this particular case, it can be noted that, as shown earlier, this approach allows the possibility of tracking SLP variations over time. It also allows the characterization of the SLP value of a material occurring at a given temperature.

In general, results show a much smaller variation of the SLP values obtained in all the peaks for device 2 when compared to the other two devices. We think the reason behind this observation is the lower temperature resolution (0.01K) of the thermal probe of device 2, as can be observed also from the temperature data in Figure \ref{fig:SAR ZigZag}a. Nevertheless, the zigzag protocol can detect this aspect, which can be taken into account when comparing results between different devices and laboratories.  

The average SLP value calculated from all the analyzed peaks in the zigzag protocol is shown in Figure \ref{fig:SAR ZigZag}d and compared with the values obtained from the classical approach that focus on the initial slope. Although devices 1 and 3 present a larger uncertainty in the SLP values obtained from the zigzag protocol in comparison with device 2, such errors are still significantly smaller than those obtained from the classical methods. Moreover, a good consistency among the obtained results from the three different devices is obtained. 

In addition, it has to be noted that the way the zigzag protocol is applied, increasing the temperature during the heating for 3 degrees and allowing to cool down for 2 degrees allows for a faster acquisition of several peaks where the data can be analyzed (as it is not necessary to wait for a complete cool down of the system).
Overall, these experimental results validate the use of the zigzag protocol to obtain reliable and reproducible SLP values independently of the device being used.

\section{Conclusion}

We have analysed the contribution to systematic deviations between measurement systems arising from the different thermal properties of the systems and sample holders. A simple 1D conduction model was used to illustrate the central factor, specifically the difference between the temperature profile during heating and cooling. This essentially arises from the fact that the heating is rather uniform, resulting in a relatively flat temperature profile in the vial during heating which becomes more strongly curved during cooling. Consequently Newton's law of cooling is not generally obeyed. However, around  the peak temperature, when the alternating field is switched off, the temperature profile remains essentially constant allowing for exact calculation of SLP by compensating for the heat losses obtained from the cooling curve. Therefore, an alternative way of calculating the SLP value was proposed, and called the Peak Analysis Method (PAM). This method was further extended into a 'zigzag protocol' by carrying out repeated heat/cool cycles. The method was verified using the simple 1D temperature diffusion model and then experimentally validated using 3 different measurement systems in different laboratories. It was shown experimentally to give significant improvement in accuracy over the 'classical' methods of SLP determination. Finally, we note that the zigzag method in principle allows the determination of a time dependent SLP as can arise, for example from the effects of chaining~\cite{Mille2021}. However, this would require more accurate measurements of heating curves. Nonetheless, the PAM technique and the zigzag protocol have shown demonstrable improvements in reproducibility between different measurement systems and laboratories and is recommended as a significant improvement over current measurement techniques.

\section{Materials and Methods section}

\subsection{Magnetic nanoparticles synthesis and characterization}
MNPs were synthesized using an oxidative precipitation aqueous route previously described  \cite{verges2008uniform} with slight modifications. Briefly, a 1 M solution of FeSO$_4$, prepared in 50 mL of H$_2$SO$_4$ 0.01 M,  was quickly added to a basic solution prepared with 4.25 g of NaNO$_3$ and 4.22 g of NaOH in a mixture of 137 mL of water and 63 mL of ethanol 96\% vol. The green rust suspension obtained was stirred for 15 min and poured in a jacketed flask previously thermalized to 90$^o$C with an thermostatic bath. The MNPs were left to grow at that temperature for 6 hours. MNPs prepared by this route were subjected to an acid treatment and then, coated with dextran \cite{teller2010magnetic}.

Size and morphology were studied by transmission electron microscopy (TEM). A drop of the diluted sample was deposited on a carbon coated grid, allowing it to dry at room temperature. Micrographs were acquired in a Tecnai G2 TEM (FEI) operated at 200 kV. The particle size was defined considering the largest internal dimension of the nanoparticles. A total of 175 nanoparticles were manually measured and the histogram obtained was fitted with a probability density function.
 
Magnetic characterization of the sample was carried out in a Quantum Design MPMS-XL SQUID magnetometer. The liquid suspension was placed in a cotton piece allowing it to dry. Then, this piece of cotton was placed in a gelatin capsule for the magnetic measurements. Field dependent magnetization of the sample was recorded at 300 K at  a maximum field of 1600 kA/m.

The magnetic hyperthermia measurements were performed using three different devices. Device 1:  commercial equipment (D5 Series from nB nanoScale Biomagnetics) with a G model closed coil. Device 2: AMF produced by a Fives Celes 12, 118 M01 generator. This device is composed of the combination of a CELES MP 6 kW generator capable of generating resonant frequencies in the range 100-400 kHz (tunable with an ALU CU type capacitor box) and a 71 mm i.d. DT25901A chilled coil. Temperature was measured with an OSENSA fiber optic probe Mod. PRB-G40-02 M-STM-MRI. Device 3: MagneTherm from Nanotherics. Two types of measurements (classical single heating-cooling cycle and a zigzag measurement ) were performed in the three devices using  a suspension of 1 ml with an iron concentration of 1mg/mL and similar AC field conditions(163.3 kHz and 35 mT for device 1, 165 kHz and 35 mT for device 2 and 172.4 kHz and 35 mT for device 3). For the classical single heating-cooling  measurements the suspension was placed into a glass vial located at the center of the magnetic induction coil inside an isolating holder. When the sample temperature was stable, the AC magnetic field was applied for 110-120 sec. The sample temperature during the heating and cooling time was recorded using a fiber optic sensor. For the zigzag heating-cooling measurements, the suspension was placed in the center of the magnetic induction coil inside an insulating support. The sample temperature was stabilized before starting the measurement. The AC magnetic field was applied until the sample temperature increased 3 $^o$C and then the AC field was switched off. The magnetic field was turned on again when the sample temperature decreased by 2 $^o$C. This process was repeated several times. The sample temperature during the heating and cooling times was recorded using a fiber optic sensor.

\subsection{Simple (1D) model of heating}
The aim is to model the increase/decrease of temperature with the SLP driving term on/off. The conjecture is as follows: the SLP is localised giving rise to a rapid (uniform) heating with losses mainly through the boundary of the sample holder, whereas during the cooling phase there will be a slow migration of heat out of the system due to small gradients within the sample holder. In 1D the time variation of the temperature (with $T=T-T_{ambient}$) is given by
\begin{equation}
\begin{split}
   \rho c  \frac{\partial T}{\partial t} &= k\frac{\partial^2T}{\partial  x^2}+S \\
   \frac{\partial T}{\partial t}&= \alpha \frac{\partial^2T}{\partial  x^2}+\frac{S}{\rho c},
\end{split}
    \label{eqn:ode}
\end{equation}
where the SLP $S$ is taken as constant, $\rho$ is the density, $c$ the heat capacity and $k$ is the thermal conductivity. $\alpha = k/(\rho C)$ is the diffusivity.
 The vial is uniformly heated with SLP $S$ for a time $t_{heat}$ after which $s$ is set to zero for the cooling phase. 
     
\subsubsection{Boundary conditions}
We now look at the experimental case of heating in a vial.  At the boundary we have to impose continuity of the temperature and the heat flux, with the interface condition
\begin{equation}
-k_{fluid}\left (\frac{dT}{dx}\right )_{fluid} = -k_{wall}\left (\frac{dT}{dx}\right )_{wall}
\label{equ:interface}
\end{equation}
In the discrete approximation, continuity of $T$ and Eqn~\eqref{equ:interface} lead to
\begin{equation}
    T_i= rT_{i-1}+T_{i+1}/(1+r),
\end{equation}
where $T_i$ is the interface temperature, $T_{i-1}$ and $T_{i+1}$ are temperatures immediately inside the fluid and boundary respectively, and the ratio $r=k_{wall}/k_{fluid}$.

 Next we consider the heat transfer to the surroundings via conduction through the vial and by convection from the upper surface of the fluid. 

The {\it convection BC }is
\begin{equation}
    -k\frac{dT}{dx} \bigg |_{x=L} = h(T_{x=L} - T_\infty),
\end{equation}
with h a constant. This leads to interface temperature at the surface of the fluid
\begin{equation}
    T_i = \frac{kT_{i-1}-h\Delta x T_\infty}{(k-\Delta x h)},
\end{equation}
where $\Delta x$ is the numerical spatial discretisation.
\subsection{The radius of curvature}
We characterise the curvature based on a radius of curvature as follows:
\begin{equation}
   r= \frac{\left ( 1+\left (\frac{d\Delta T}{dx}\right )^2 \right )^{1.5}}{\frac{d^2\Delta T}{dx^2}}
   \label{eq:radius}
   \end{equation}
The "-" sign, indicates the concave temperature profile and the absolute value quantifies how large the non-uniformity is, with zero meaning constant temperature. Note that because the initial radius of curvature at time $t=0$ is infinite we chose to use a measure of curvature as $r^{-1}$.

\begin{acknowledgement}
Authors would like to acknowledge financial support from the Ministerio de Ciencia,Innovaci\'on y Universidades (MCIU), the Agencia Estatal de Investigaci\'on (AEI), and Fondo Europeo de Desarrollo Regional (FEDER) through projects: PGC2018-096016-B-I00 (to L.G.), PID2020-113480RB-I00 (to S.V. and P.M.) and PID2019-109514RJ-100 (to D.S.).  Xunta de Galicia is acknowledged for project ED431F 2022/005 (to D.S.). AEI is also acknowledged for  the (\textit{Ram\'on y Cajal} grant RYC2020-029822-I to D.S. We acknowledge the Centro de Supercomputacion de Galicia (CESGA) for computational resources. 
\end{acknowledgement}

\bibliography{sample}

\end{document}


\begin{suppinfo}

\section{ Section 1. Magnetic nanoparticles synthesis and characterization details}

Iron oxide magnetic nanoparticles were synthesized by precipitation of an iron (II) salt ($FeSO_{4}$) in the presence of a base (NaOH) and a mild oxidant ($KNO_{3}$) following a modified version of the synthesis described by Verges et al.\cite{verges2008uniform}. The obtained particles (32.0$\pm$6.7 nm of average size, see Figure \ref{fig:NPs Characterization}a) were then coated with dextran to improve their stability in water. The field-dependent magnetization of the particles showed saturation magnetization, remanent magnetization and coercive field values of $\approx$ 77 Am$^2$/kg$_{Fe_3O_4}$, $\approx$ 12 Am$^2$/kg$_{Fe_3O_4}$ and $\approx$ 4 kA/m, respectively (Figure \ref{fig:NPs Characterization}b).

\begin{figure}[bth!]
\centering
\includegraphics[width=1.0\linewidth]{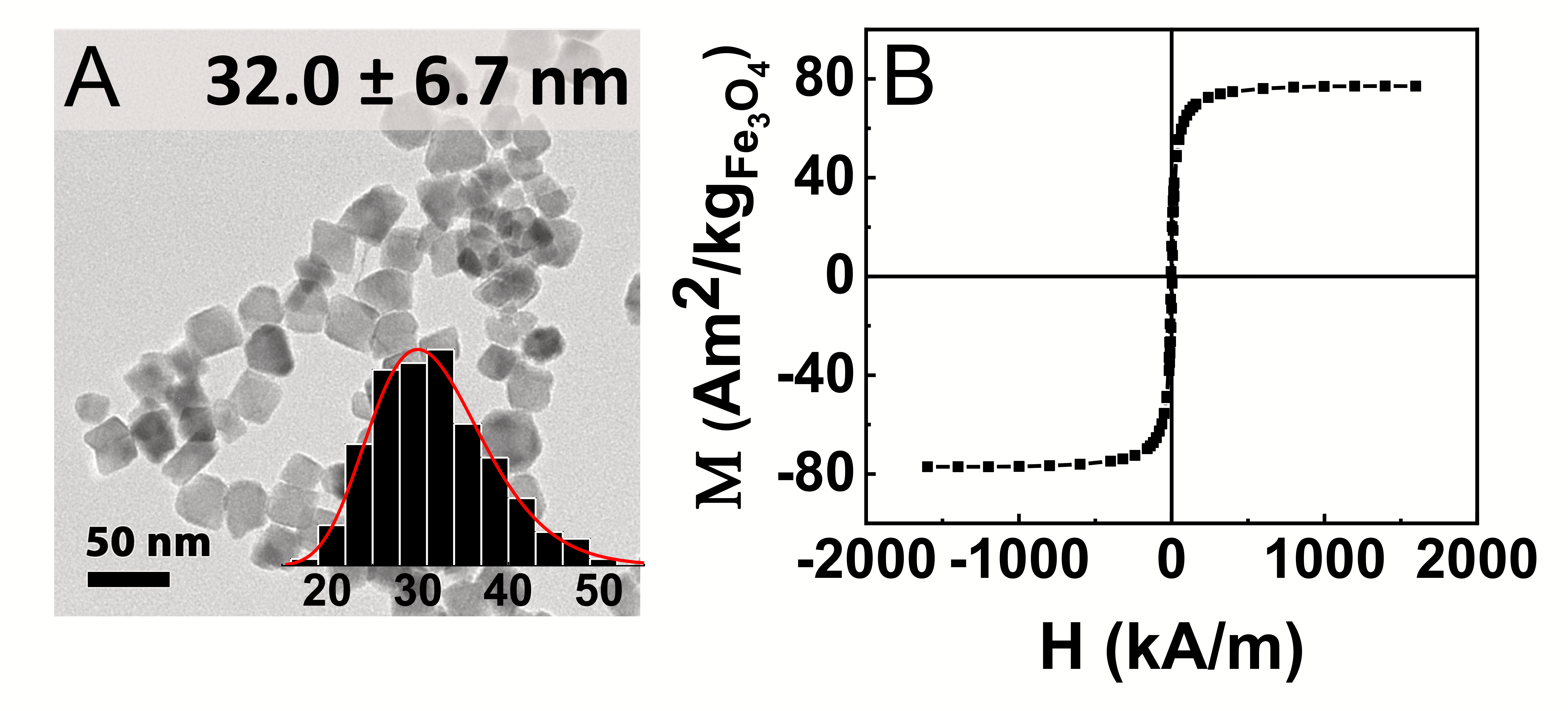} 
\caption{Magnetic nanoparticles characterization.(A) Transmission electron microscopy image and particle size distribution histogram for nanoparticles . (B) Field dependent magnetization at room temperature (300 K) .}
\label{fig:NPs Characterization}
\end{figure}

\section{Section 2. Fit of the initial slope with the classical methods}

SLP values were calculated using the ISM, BLM and CSM for the analysis of the initial slope of the calorimetric results (Figure 1 from the manuscript). However, when looking in detail at the first minute of the heating curve, some differences were found among the devices. Probably associated to the lower resolution of the temperature probe of device 1, measurements performed with this device presented some "jumps" in the obtained data. A delay on the onset of the temperature increase was also observed in this case. This led to some problems associated to the calculation of the SLP value using the Initial Slope, the Box-Lucas  or the Corrected Slope Methods. In particular, in the case of device 1, the Box Lucas iteration did not converge and the "L" value from the corrected slope method had a nonphysical value. These results indicated that the quality of the acquired data had a strong impact on the uncertainties associated to the SLP value calculation. Data obtained from devices 2 and 3, were smoother and did not present any problem in the fit using the three different approaches.

\newpage

\begin{table}[H]
\rowcolors{3}{gray!25!}{gray!10!}
\footnotesize
\centering
\caption{Fit parameters}
  \begin{tabular}{|c|c|c|c|c|c|c|c|c|c|} 
 \hline
   & & \multicolumn{2}{|c|}{\textbf{ISM}}& \multicolumn{4}{|c|}{\textbf{BLM}}& \multicolumn{2}{|c|}{\textbf{CSM}} \\
\hline
  \textbf{Devices} & \textbf{Measurement} & \textbf{Slope  (K/s)} & \textbf{SLP (W/g)} & \textbf{b(1/s)} & \textbf{a(K)} & \textbf{L(W/K)} & \textbf{SLP (W/g)} & \textbf{L(W/K)}  & \textbf{SLP (W/g)} \\ [0.5ex] 
 \hline
  Device 1    &     1  &  0.075  &  230  &	-0.01  &	-9.13  & -0.030  &	201  & -0.044  &	194\\
   \hline
  Device 1    &     2  &  0.095  &  287  &	$<$0.001  &	-192.83  &	-0.002  &	293  & 0.012  &	271\\
\hline
  Device 2    &     1  &  0.084  &  254  &	0.01  &	9.68  &	0.042  &	291  & 0.044  &	295\\
   \hline
  Device 2    &     2  &  0.078  &  234  &	0.01  &	12.62  &	0.028  &	258  & 0.020  &	252\\
\hline
  Device 3    &     1  &  0.081   &	246  &	$<$0.001  &	5472.70  &	$<$0.001 & 230  &	0.005 &  235\\
   \hline
  Device 3    &     2  &  0.078  &  237  &	0.004 &	18.26  & 0.019  &	246  &  0.024  &	231\\
\hline
\end{tabular}
\label{tabla:Informacion nanoparticulas}
\renewcommand{\tablename}{Tabla}
\end{table}

To fit the cooling part of figure 3 within the main text, the first approach was to impose the condition of continuity in the temperature vs. time curve (i.e. the initial T of the cooling part is the final one reached during the heating part), it is obtained $a=0.0020$, a rather poor fit, characterised by $R^2=0.972$, and also directly seen by the scarce overlapping between the fitting curve (dotted blue line) and experimental data (thick solid grey line). An improvement in the fit is reached is the initial condition is not applied, so that $R^2=0.992$, corresponding to $a=0.0017$. Furthermore, an even greater fit (characterised by a higher $R^2$ value) can be obtained if, in addition to removing the condition of initial temperature, the fit is restricted to the final part of the cooling curve. In this case (green solid line inf figure 3), it is obtained $a=0.0015$, with $R^2=0.993$. The reason why in this range the single-exponent fitting works better is because in this range the lossed do behave more in a single heat-loss channel manner, as assumed in Newton's law. This is clearly illustrated in the inset of figure 3, as the final part of the curve corresponds to a linear behaviour of the $\frac{d(\Delta{T})}{dt}$ \textit{vs.} $\Delta{T}$ data.)

\section{Section 3. Parameters for the heat diffusion calculations}

\begin{table}[H]
\centering
\begin{tabular}{|l|l|l|}
\hline 
1.d3    &      $ \rho_{vess}$ & $kg/m^3$\\ \hline 
1.d3    &      $ \rho_{wall}$ & $kg/m^3$ \\ \hline 
4.2d3   &      $c_{Vvess}$    & $J/(kg K)$ \\ \hline 
4.2d3   &      $c_{Vwall}$    & $J/(kg K)$ \\ \hline 
1000    &          SAR        & $W/g$ \\ \hline 
0.8d0   &      $k_{vess}$ & $Wm^{-1}K^{-1}$ \\ \hline 
0.8d0   &      $k_{wall}$ & $Wm^{-1}K^{-1}$  \\ \hline 
\end{tabular}
\caption{Parameters for the heat diffusion calculations.}
\label{tab:parameters1}
\end{table}

\section{Section 4. Standard Operating Procedure}

A Standard Operating Procedure to perform the Zig-zag protocol is described below.
\begin{enumerate}
\item Place the MNPs liquid suspension in the sample holder (ideally  1 mL volume at a concentration of 1 mg Fe/mL for the case of iron oxide magnetic nanoparticles).
\item Wait until the temperature is stable before turning the AC magnetic field on.
\item Turn the AC field ON and wait until the sample temperature has increased 3ºC. 
\item Switch OFF the AC field and wait until the temperature has decreased 2ºC.
\item Turn the field ON again and repeat the cycle of heating 3ºC and cooling 2ºC around 15 times.
\item Calculate the slope of the heating and cooling curves near the peak (although this may depend on the device used for the measurements, a possible interval for this could be around 0.5 ºC).
\item Following Equation 8 from the manuscript, the SLP is calculated from the difference of the two slopes. 
\end{enumerate}

\bibliography{sample}

\end{suppinfo}